\documentclass[conference]{IEEEtran}
\usepackage{cite}
\usepackage{amsmath,amssymb,amsfonts}
\usepackage{graphicx}
\usepackage{textcomp}
\usepackage{xcolor}
\usepackage{qcircuit}
\usepackage{caption}
\usepackage{subcaption}
\usepackage{algorithm}
\usepackage{algpseudocode}

\def\BibTeX{{\rm B\kern-.05em{\sc i\kern-.025em b}\kern-.08em
    T\kern-.1667em\lower.7ex\hbox{E}\kern-.125emX}}

\newcounter{daggerfootnote}

\newcommand*{\daggerfootnote}[1]{%
    \setcounter{daggerfootnote}{\value{footnote}}%
    \renewcommand*{\thefootnote}{\fnsymbol{footnote}}%
    \footnote[2]{#1}%
    \setcounter{footnote}{\value{daggerfootnote}}%
    \renewcommand*{\thefootnote}{\arabic{footnote}}%
    }
\begin{document}

\title{QAOA of the Highest Order}

\author{\IEEEauthorblockN{Colin Campbell}
\IEEEauthorblockA{\textit{Applications Team} \\
\textit{ColdQuanta}\\
Boulder, Colorado \\
colin.campbell@coldquanta.com}
\and
\IEEEauthorblockN{Edward Dahl}
\IEEEauthorblockA{\textit{Applications Team} \\
\textit{ColdQuanta}\\
Boulder, Colorado \\
denny.dahl@coldquanta.com}
}

\maketitle

\begin{abstract}
The Quantum Approximate Optimization Algorithm (QAOA) has been one of the leading candidates for near-term quantum advantage in gate-model quantum computers. From its inception, this algorithm has sparked the desire for comparison between gate-model and annealing platforms. When preparing problem statements for these algorithms, the predominant inclination has been to formulate a quadratic Hamiltonian. This paper gives an example of a graph coloring problem that, depending on its variable encoding scheme, optionally admits higher order terms. This paper presents evidence that the higher order formulation is preferable to two other encoding schemes. The evidence then motivates an analysis of the scaling behavior of QAOA in this higher order formulation for an ensemble of graph coloring problems.
\end{abstract}


\section{Introduction}
For gate-model quantum computers and annealing platforms, QAOA and the Quantum Adiabatic Algorithm (QAA), respectively, are positioned to solve the same type of problem—quadratic Ising model, or equivalently, Quadratic Unconstrained Binary Optimization (QUBO) \cite{quboref}. These problems take the form of equations (1) and (2). Although the problems are equivalent up to a change of variable, since QAOA naturally uses the Ising basis, this paper will mainly represent binary variables as $s \in \{1, -1\}$.
\begin{equation}
    Q = \sum_{i}w_{i}x_{i} + \sum_{i<j}w_{ij}x_{i}x_{j} \quad x_{i} \in \{ 0, 1 \}
\end{equation}
\begin{equation}
    H = \sum_{i}h_{i}s_{i} + \sum_{i<j}h_{ij}s_{i}s_{j} \quad s_{i} \in \{ 1, -1 \}
\end{equation}

These problems represent NP-Complete decision problems with NP-Hard optimization variants \cite{arora}. Many significant industrial applications can be formulated in this way \cite{glover, ajinkya, hodson, tobias}, driving much of the interest in these algorithms as opportunities to demonstrate near-term quantum business advantage. Projections for industrial relevance vary, but some projections put QAOA style advantage particularly soon \cite{qutac, zhou}. To accelerate this development as much as possible, using all the tools available to gate-model systems will be necessary. At the time of writing, there appear to be only a handful of examples utilizing higher order terms (HOTs) in QAOA\cite{hadfield, hungary_poland, fakhimi, glos, fuchs}. The four corners map coloring problem is a small, hand-picked member of a larger ensemble of NP-Complete coloring problems that could benefit from this formulation.

 When formulating combinatorial optimization problems to assess QAOA or QAA, the focus appears to have been on problems naturally admitting quadratic objectives \cite{lucas}. On annealers, this inclination is well-justified because all commercial quantum annealing platforms require at most quadratic interactions for the final computation \cite{dwave}. While order reduction is possible, such techniques necessarily introduce more variables and interactions, even if at only polynomial cost \cite{boros, mandal}. QAOA does not have this inherent limitation because gate-model machines can naturally implement HOTs, effectively expanding the problem space to include products of more than two binary variables. The circuit representation of these HOTs is a subset of a family of operators called Pauli Gadgets that represent exponentiated tensor products of Pauli matrices. Figures 1 and 2 show possible circuit decompositions of a quartic term into single-qubit rotations and CX gates. For more information on Pauli Gadgets and their decompositions see \cite{phasegadgets}.
 
 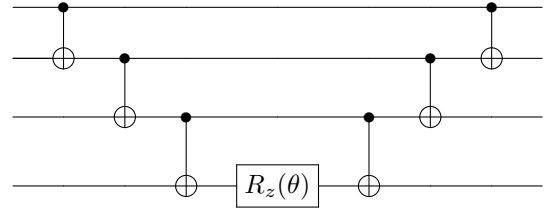
\begin{figure}[htbp]
        \centerline{
         \Qcircuit @C=1.5em @R=1.4em {
            & \ctrl{1} & \qw & \qw & \qw & \qw & \qw & \ctrl{1} & \qw\\
            & \targ \qw & \ctrl{1} & \qw & \qw & \qw & \ctrl{1} & \targ & \qw \\
            & \qw & \targ & \ctrl{1} & \qw & \ctrl{1} & \targ & \qw & \qw \\
            & \qw & \qw & \targ & \gate{R_{z}(\theta)} & \targ & \qw & \qw & \qw
            }
        }
\caption{A Decomposition of $\exp\left(-i\frac{\theta}{2}Z\otimes Z \otimes Z \otimes Z\right)$} 
\label{fig}
\end{figure}

 \begin{figure}[htbp]
        \centerline{
         \Qcircuit @C=1.5em @R=1.4em {
            & \ctrl{1} & \qw & \qw & \qw & \ctrl{1} & \qw\\
            & \targ \qw & \targ & \gate{R_{z}(\theta)} & \targ & \targ & \qw \\
            & \targ & \ctrl{-1} & \qw & \ctrl{-1} & \targ & \qw \\
            & \ctrl{-1} & \qw & \qw & \qw & \ctrl{-1} & \qw
            }
        }
\caption{Another Decomposition of $\exp\left(-i\frac{\theta}{2}Z\otimes Z \otimes Z \otimes Z\right)$} 
\label{fig}
\end{figure}
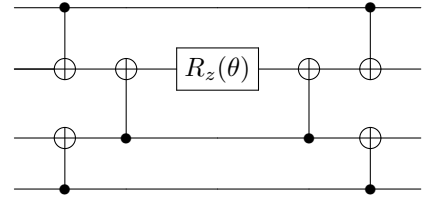

With access to this direct circuit representation of HOTs, there is apparent need to expand potential near-term applications of QAOA to include problems that naturally admit HOTs. While much of the gate-model community has focused on MAXCUT \cite{farhi, wang, wurtz, googleai, ruslan}, and the annealing community has considered a richer ensemble of natively quadratic problems\cite{grant, neukart}, higher order Ising formulations offer an opportunity to enrich the applications space for QAOA. By taking inspiration from the annealing community’s mature space of formulating QUBOs and incorporating a unique tool of gate-model platforms, this paper poses the four corners map coloring problem as an example benefiting from HOTs.

Section II poses the four corners problem as an Ising Model using three different encoding schemes: the natively higher order binary encoding, the order reduced binary encoding, and the natively quadratic unary encoding. Section III compares the performance of QAOA across all three. Section IV compares the scaling behavior of the binary and unary encoding schemes and presents two optimizations available to the binary case that are only possible due to the inclusion of HOTs. Lastly, Section V draws conclusions and suggests topics for future work.

\section{Four Corners Map Coloring Problem}
The four corners map coloring problem (Figure 3) is a simple problem belonging to a larger ensemble of graph coloring problems. Section V presents the more general statement of the family of problems, but the specific instance considered first provides a concrete example to motivate the general characterization. The goal of this problem is to, with access to four colors, find a coloring of the four corners map such that no two adjacent regions have the same color (not counting diagonal regions as adjacent).

\begin{figure}[htbp]
\centerline{\includegraphics[width=0.5\textwidth]{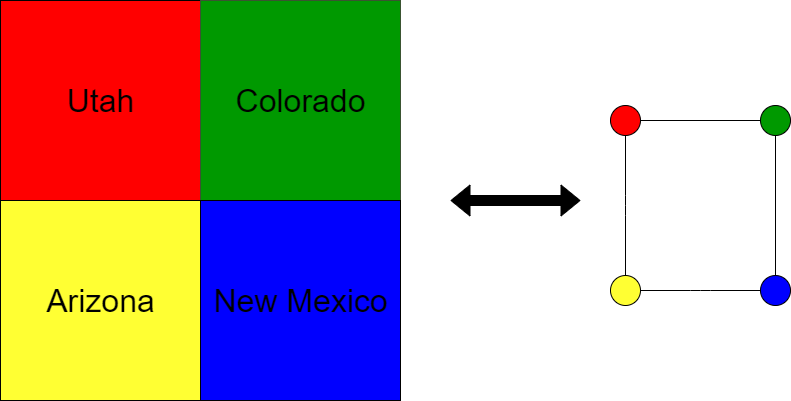}}
\caption{One valid coloring of the four corners map with binary spin vector $(1,1,-1,1,1,-1,-1,-1)$}
\label{fig}
\end{figure}

\subsection{Binary Encoding}
This graph coloring problem admits an eight variable Ising formulation by using a binary encoding scheme. The binary encoding scheme encodes the color of a node in terms of spin variables corresponding to its binary expansion. For now, the eight variables are labeled as $( a_{0}, a_{1}, b_{0}, b_{1}, c_{0}, c_{1}, d_{0}, d_{1} )$, where the letter corresponds to the region on the map (clockwise starting with Utah), and the subscript corresponds to the bit position of the color's binary expansion. The spin representations of the four colors red, blue, green, and yellow are (+1+1), (+1-1), (-1+1), (-1-1), respectively. Therefore, for node $a$ two variables $a_{0}$ and $a_{1}$ represent all four possible colors.

The simplest way to generate the penalty for two adjacent nodes $a$, $b$ having the same color is to note that if the undesired condition is true, all of the bits representing that color will be the same. Therefore, we use the following expression:
\begin{equation}
    (1+a_{0}b_{0})(1+a_{1}b_{1}),
\end{equation}
which is zero if $a_{i} \neq b_{i}$ and positive otherwise.

Since (3) is minimized when Utah and Colorado are different colors,  the whole problem may be represented as a sum over the map's borders, using (3) with appropriate variables. With the given variable assignments, this process produces:
\begin{align}
    H_1= & \ \ \ (a_{0}a_{1} + c_{0}c_{1})(b_{0}b_{1} + d_{0}d_{1}) \\ \nonumber 
    & + (a_{0} + c_{0})(b_{0} + d_{0}) + (a_{1} + c_{1})(b_{1} + d_{1}) + 4.
\end{align}
This quartic Ising model is equal to zero if and only if the spin variable assignment $( a_{0}, a_{1}, b_{0}, b_{1}, c_{0}, c_{1}, d_{0}, d_{1} )$ represents a valid coloring of the four corners graph. In total, the simplest circuit representation of $\exp (-i\gamma H_{1})$ in QAOA requires 40 two-qubit gates, a depth which will prove significantly better than the other encodings even without further optimization.

\subsection{Order Reduction}
If a quadratic objective were strictly required, order reduction can remap the quartic Ising Model to a quadratic one. Order reduction proceeds by replacing the product of a pair of variables with a new expression that is at most linear in the existing variables and a single auxiliary variable.  All terms containing the product can be rewritten with the replacement rule, reducing the overall order by one.  To ensure that the product of the original pair of variables is equal to its replacement expression, there must be an additional set of terms which is minimized exactly when equality holds.  For example, to replace the variable pair $a_0 a_1$, the replacement expression containing the auxiliary variable $A$ has the following replacement rule (5) and corresponding constraint (6):
\begin{align}
    a_0 a_1 = 1 + a_0 + a_1 - 2A\\
    -2A(a_0 + a_1 + 1) + (1 + a_0)(1 + a_1) + 2.
\end{align}
Verifying that (5) holds when (6) is minimized is straightforward and reduces the order of the expression by one.

Until the overall order is quadratic or less, repeat this process.  Each replacement requires a new auxiliary variable, doubling the binary search space.  Further complicating the process, if the penalty incurred by violating the equality constraint is less than the savings gained in other parts of the Hamiltonian, applying a positive weight to the constraint terms, called a Lagrange Parameter, will be necessary\cite{boros}. 

Substituting (5) into (4) and collecting the constraint terms (6) yields:
\begin{align}
    H_{obj} =& \ (2+a_0+a_1+c_0+c_1-2A-2C) \\ \nonumber
    & \times (2+b_0+b_1+d_0+d_1-2B-2D) \\ \nonumber
    & + (a_0+c_0)(b_0+d_0)+(a_1+c_1)(b_1+d_1) + 4, \\ \nonumber \\
    H_{con} =& -2A(a_0+a_1+1)+(1+a_0)(1+a_1) \\ \nonumber
    & -2B(b_0+b_1+1)+(1+b_0)(1+b_1) \\ \nonumber
    & -2C(c_0+c_1+1)+(1+c_0)(1+c_1) \\ \nonumber
    & -2D(d_0+d_1+1)+(1+d_0)(1+d_1) + 8.
    \end{align}
Therefore, the final order-reduced Hamiltonian becomes:
\begin{equation}
    H_2 = H_{obj} + \lambda H_{con},
\end{equation}
where choosing $\lambda \geq 3$ ensures that breaking the constraints is never favorable. After order reduction, this encoding scheme seems a poor choice because of the introduction of four more variables, many interaction terms, and an extra sub-problem to encourage the new variables to behave properly. In total, the reduction yields 48 quadratic terms corresponding to 96 CX gates to implement $\exp\left(-i\gamma H_{2}\right)$ with no hope of further optimization. If a quadratic Ising formulation is strictly required, an alternative variable encoding scheme with less overhead is desirable.

\subsection{Unary Encoding}
The Unary encoding scheme is a form of one-hot encoding in the binary spin variables. Once again, letters will indicate regions, but this time the subscripts $i=0, 1, 2, 3$ will directly represent the colors red, blue, green, and yellow, respectively. For example, $c_{1}=-1$ if New Mexico is assigned blue and $c_{1}=1$ otherwise. With this variable encoding scheme, the term penalizing two adjacent nodes $a,b$ having the same color is given by:
\begin{equation}
    \sum_{i=0}^{3}(1-a_{i})(1-b_{i}).
\end{equation}

However, an additional set of constraints is required to ensure only one color is chosen within a region. For example, the constraint that Utah must be one color is a 1-of-4 constraint of the form:
\begin{equation}
    \left( -2 + \sum_{i=0}^{3}a_{i} \right)^{2}.
\end{equation}
To get the full Ising model representation in the unary encoding scheme, both sets of constraints must be satisfied, so the final objective is the sum of the two sub-problems:
\begin{equation}
    H_3 = \sum_{i=0}^{3}(2 - a_{i} - c_{i})(2 - b_{i} - d_{i}) + \sum_{s} \left( -2 + \sum_{i=0}^{3}s_{i} \right)^{2},
\end{equation}
where $s  \in \{a, b, c, d\}$ is the letter representing each region. This objective has 40 quadratic terms, represented by 80 two-qubit gates also without possibility of further optimization.

\section{Comparing Performance}
While the raw number of gates gives a good indication of performance expectations for each of the three encodings, some energy landscapes could still be more appealing than others. To compare their impact on QAOA, 10 iterations of Constrained Optimization by Linear Approximation (COBYLA) in a noise-free simulation of the evolution of the statevector up to $p=5$ provided samples\protect\daggerfootnote{14 out of the 150 total samples were determined to be outliers and removed from the plot because they fell outside 1.5 times the interquartile range.} for comparison. The results are displayed in Figure 5. Other optimizers were considered, but COBYLA proved the most consistent at small $p$. For each run the relative error and the total probability of measuring the winning state were recorded. While the success probability favors the lower variable encoding due to the exponential binary search space, the ultimate goal of QAOA is measuring a valid solution with high probability. Reporting this probability is therefore appropriate for comparative analysis. Notably, despite using the most variables unary encoding scheme performs no worse than the binary reduction with respect to this metric.

To gain intuition into what makes these encoding schemes better or worse, Figure 4 shows the $(\beta, \gamma)$ plots of the relative error for $p=1$ grid search done with $128 \times 128$ points. From the plots, traversing any of these spaces can result in finding poor local minima, often referred to as "Barren Plateaus,"\cite{pesah} but the binary encoding scheme provides the best opportunity to find a high quality solution, at least for $p=1$.

When considering higher $p$, the binary encoding scheme still performs best. In fact, from the simulations, a clear correlation between increasing $p$ and significantly increasing the success probability is not apparent. This result is unsurprising because, in the case of the binary encoding scheme, the results are already high quality at $p=1$. For the binary reduction and unary encodings, QAOA is performing so poorly that both a higher $p$ value and a different classical optimizer are likely necessary to see improvement. However, simulations for high $p$ become computationally intensive and outside the scope of this work, especially considering QAOA can solve the problem well for low $p$ and a standard optimizer. Nevertheless, the results of the different encoding schemes for the four corners map coloring problem make abundantly clear the fact that using higher order terms to represent the problem Hamiltonian has considerable benefits on the performance of QAOA compared to other encoding schemes that require quadratic terms as the highest order.

\begin{figure}[t]
    \centering
    \includegraphics[width=0.5\textwidth]{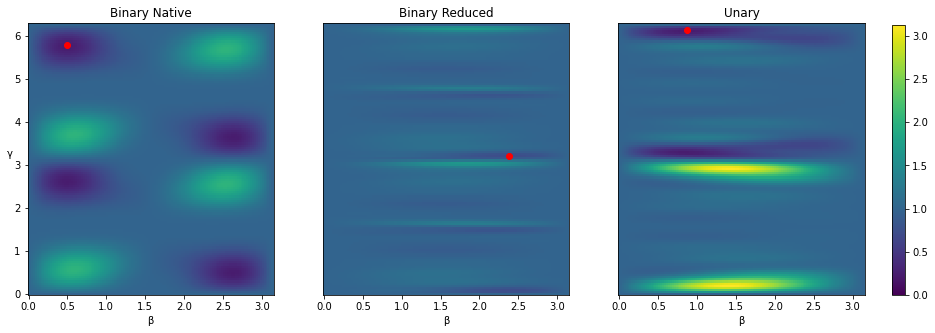}
    \caption{These plots show the $(\beta, \gamma)$ variation of relative error from the true energy minimum at $p=1$ with the best point marked in red. These plots give a strong indication that the binary encoding scheme has both higher quality minima and a higher chance to end in a high quality solution.}
    \label{fig:my_label}
\end{figure}

\section{General Scaling Behavior}
Since the order reduction is not a strict requirement for gate model systems and appears to perform worse than its higher order counterpart, this section compares the scaling behavior of the unary encoding to the native binary encoding scheme on the general graph coloring problem.

Given a Graph $G=(V,E)$ and a set of colors $C=\{0,...,c-1\}$, the general graph coloring problem asks to find an assignment $f: V \to C$ such that if $(v_{i}, v_{j}) \in E$ then $f(v_{i}) \neq f(v_{j})$. Let $n=|V|, e=|E|, c=|C|$ be the number of vertices, edges, and colors, respectively. To build the quadratic Ising model in the unary encoding scheme, let $\mu_{i}^{k} = -1$ if $f(v_{i}) = c_{k}$ and $1$ else. As before, there must be two types of penalty terms:
\begin{align}
    H_{con}=& \sum_{i=0}^{n-1} \left(2-c + \sum_{k=0}^{c-1}\mu_{i}^{k} \right)^{2}\\
    H_{adj}=& \sum_{(v_{i},v_{j}) \in E} \left(\sum_{k=0}^{c-1}(1 - \mu_{i}^{k})(1 - \mu_{j}^{k}) \right).
\end{align}

Counting the quadratic terms in (13) and (14) determines the exact quantum resources needed to represent the problem Hamiltonian in QAOA. The number of qubits needed to represent the general unary representation is $O(nc)$. The number of quadratic terms in the constraint term is $\frac{nc(c-1)}{2}$, and the number of quadratic terms in the adjacency term is $ce$. Because the number of two-qubit gates is twice the number of quadratic terms, the total number of two-qubit gates is $2ce + nc(c-1)$. If $e \propto n$, the total number of two-qubit gates scales as $O(nc^{2})$, with the regional constraint dominating.

\begin{figure}[t]
    \centering
    \includegraphics[width=0.5\textwidth]{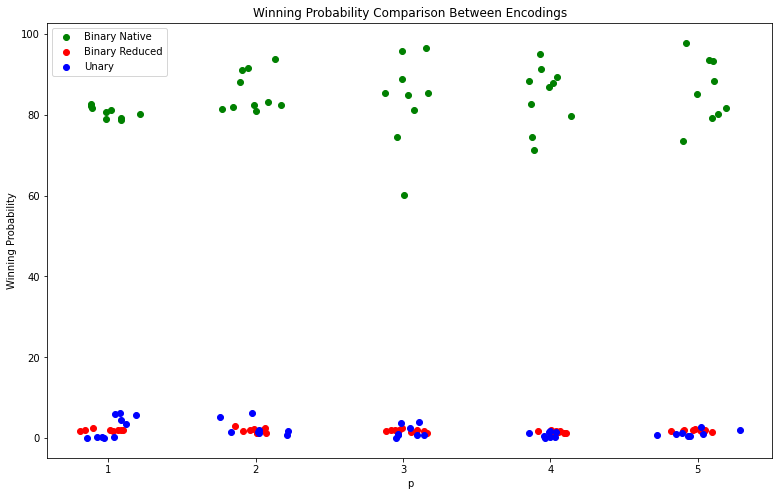}
    \caption{This graphic shows the winning probability of 10 samples of COBYLA optimization for each encoding scheme up to $p=5$, where the x-coordinate has been jittered slightly away from its $p$ value to make the variation easier to see.}
    \label{fig:my_label}
\end{figure}

The Binary encoding scheme yields better scaling behavior in both space and depth, especially in the case where $c = 2^{l}$ for some $l \in \mathbb{Z}_{\geq 0}$. Let $\sigma_{i}^{k} = -1$ if the $k$th digit of the binary expansion of $f(v_{i}) \in C$ is $1$ and $1$ else, and let $l = \lceil \log_{2}(c) \rceil$. For example, consider the case in which $f(v_{3})=12$. Since the binary representation of $12$ is $1100$, $\sigma_{3}^{0}=\sigma_{3}^{1}=1$ and $\sigma_{3}^{2}=\sigma_{3}^{3}=-1$. Therefore the number of qubits is exactly $nl$, and the spatial resource scaling is $O(n\lceil\log_{2}(c)\rceil)$.

To determine the depth, first consider the general formula for the adjacency constraint in the binary encoding scheme:
\begin{align}
    H_{adj}=& \sum_{(v_{i}, v_{j}) \in E} \left(\prod_{k=0}^{l-1}(1 +\sigma_{i}^{k}\sigma_{j}^{k})\right).
\end{align}
When $c$ is exactly a power of two, this adjacency term is sufficient, but if $c$ is not a power of two, there must be penalty functions ensuring certain spin states are illegal, effectively constraining the number of valid colors to $c$ at any given node. In the absolute worst case, where $c$ is one less than a power of two, there must a single state penalty function on each node $v_{i}$ with no possibility of cancellation within its terms. Summing this penalty over all nodes yields the total regional constraint:
\begin{align}
    H_{con}=& \sum_{i=0}^{n-1} \left(\prod_{k=0}^{l-1}(1 + \sigma_{i}^{k})\right).
\end{align}

The total number of gates from the sum of the gates in (15) and (16) can be determined from the order of terms in the expression and the number of unique terms of each order. Specifically, a term of order $m \geq 2$ requires $2(m-1)$ two-qubit gates, as in Figures (1) and (2). Algebraically, the total number of gates can be written as:
\begin{align}
    2e\sum_{k=1}^{l-1}{l \choose k}(2k-1) + 2n\sum_{k=1}^{l-1}{l \choose k}(k-1).
\end{align}
After simplification, the total becomes:
\begin{equation}
    \left(2^{l}-2\right) \left[ (2e+n)l-2(e+n) \right].
\end{equation}
Once again letting $e \propto n$ determines the total number of two-qubit gates to scale as $O(nc\lceil\log(c)\rceil)$.

Despite both the space and depth resources being superior in this encoding scheme, the binary formulation still has opportunities to perform significantly better than the worst case would suggest, completely unlike the unary case. There exist at least two strategies to reduce depth. First, Algorithm\protect\daggerfootnote{bin$(x)$ returns the bit string of $x$ starting from the most significant bit.} 1 avoids the worst case scenario whenever possible by generating multi-state penalties (MSPs) to exclude an exponential number of spin states at each step. The algorithm only needs to run once for some node $j$ and can then be reproduced for any other node by replacing $j$ with the correct node label. The second strategy involves composing sub-circuits like those in Figures 1 and 2 in a way that produces cancellation within the logical circuit. For example, cancellation of 16 CX gates is possible in the four corners problem. Detailing the circuit representation that produces the most cancellation is out of scope of this paper, but interested readers can refer to \cite{seyon} for more details.

\begin{algorithm}
\caption{Penalize Illegal Spin States on Node $j$}
\textbf{Input: $c \geq 1$}
\begin{algorithmic}[1]
\State $l \gets\lceil \log_{2}(c) \rceil$
\State $p \gets 2^{l} - c$
\Comment{\textit{number of states to penalize}} 
\State $t \gets 2^{l} - 1$
\Comment{\textit{next state to penalize}}
\State $R \gets 0$
\Comment{\textit{penalty function to return}}
\While{$p > 0$}
    \State $b \gets$ bin$(t)$
    \State $m \gets l - \lfloor \log_{2}(p) \rfloor$
    \Comment{\textit{number of free bits}}
    \State $P \gets 1$
    \Comment{\textit{product to generate MSP}}
    \For{$k = m:l-1$}
        \If{$b_{k}$ is $0$}
            \State $P \gets P \times (1 + \sigma_{j}^{k})$
        \Else
            \State $P \gets P \times (1 - \sigma_{j}^{k})$
        \EndIf
    \EndFor
    \State $R \gets R + P$
    \Comment{\textit{add MSP to total penalty}}
    \State $t \gets t - 2^{l-m}$
    \State $p \gets p - 2^{l-m}$
\EndWhile
\State \Return $R$
\end{algorithmic}
\end{algorithm}

\section{Conclusion}
This paper explores the four corners map coloring problem to provide evidence suggesting that utilizing HOTs can significantly improve the performance of QAOA. Building on this intuition leads to the general scaling behavior of this formulation compared to an alternate encoding, further demonstrating that the higher order Ising formulation can confer benefits beyond the specifically chosen example. More investigation is necessary to determine the efficacy of the difference in complexity scaling for larger graphs, where simulation quickly becomes unfeasible. As NISQ devices continue to improve in both qubit count and fidelity, the authors look forward to directly comparing QAOA's performance with different encoding schemes in the hopes of accelerating the time to quantum advantage on large ensembles of NP problems that admit HOTs.


\begin{thebibliography}{00}
\bibitem{quboref} G. Kochenberger et al, ``The unconstrained binary quadratic programming problem: a survey'', J. Comb. Optim., pp. 58-81, vol. 28, 2014.

\bibitem{arora} S. Arora, B. Barak, ``Computational Complexity,'' Cambridge University Press, 2009, pp. 38-67.

\bibitem{glover} F. Glover, G. Kochenberger, ``A Tutorial on Formulating QUBO Models'', Computing Research Repository, 2018, \textit{arXiv:1811.11538}.
\bibitem{ajinkya} A. Borle, V. Elfing, S. Lomonaco "Quantum Approximate Optimization for Hard Problems in Linear Algebra", 2020, \textit{arXiv:2006.15438}.
\bibitem{hodson} M. Hodson et al, "Portfolio rebalancing experiments using the Quantum Alternating Operator Ansatz," 2019, \textit{arXiv:1911.05296v1}.
\bibitem{tobias} T. Stollenwerk, S. Hadfield and Z. Wang, "Toward Quantum Gate-Model Heuristics for Real-World Planning Problems," in IEEE Transactions on Quantum Engineering, vol. 1, pp. 1-16, 2020, Art no. 3101816, doi: 10.1109/TQE.2020.3030609.

\bibitem{qutac} Quantum Technology and Application Consortium – QUTAC., Bayerstadler, A., Becquin, G. et al. Industry quantum computing applications. EPJ Quantum Technol. 8, 25 (2021). https://doi.org/10.1140/epjqt/s40507-021-00114-x.
\bibitem{zhou}L. Zhou et al, "Quantum Approximate Optimization Algorithm: Performance, Mechanism, and Implementation on Near-Term Devices", Phys. Rev. X 10, 021067, 2020, doi: 10.1103/PhysRevX.10.021067.

\bibitem{hungary_poland} A. Glos, A. Krawiec, Z. Zimboras, ``Space-efficient binary optimization for variational computing'', 2020, \textit{arXiv:2009.07309v1}.
\bibitem{glos} Z. Tabi et al, "Quantum Optimization for the Graph Coloring Problem with Space-Efficient Embedding," 2020, \textit{arXiv:2009.07314}.
\bibitem{fuchs}Fuchs, F.G., Kolden, H.Ø., Aase, N.H. et al. Efficient Encoding of the Weighted MAX k-CUT on a Quantum Computer Using QAOA. SN COMPUT. SCI. 2, 89 (2021). https://doi.org/10.1007/s42979-020-00437-z
\bibitem{hadfield} S. Hadfield, "On the representation of Boolean and real functions as Hamiltonians for quantum computing," 2018, \textit{arXiv:1804.09130}.
\bibitem{fakhimi} R. Fakhimi et al, "Quantum-inspired Formulations for the Max \textit{k}-cut
Problem," ISE Technical Report 21T-007 Lehigh University.

\bibitem{lucas} A. Lucas, "Ising formulations of many NP problems," Front. Phys., vol. 2, 12 February 2014, doi: 10.3389/fphy.2014.00005.

\bibitem{dwave} C. McGeoch, R. Harris, S. Reinhardt, P. Bunyk, ``Practical Annealing-Based Quantum Computing'', Computer. 52. p. 38-46, 2019.

\bibitem{boros} E. Boros, P. Hammer, Discrete Applied Mathematics, vol. 123, pp. 155-225, November 2002.
\bibitem{mandal}A. Mandal, "Compressed Quadratization of Higher Order Binary Optimization Problems," 2020, \textit{arXiv:2001.00658}.

\bibitem{phasegadgets} A. Cowtan et al, "Phase Gadget Synthesis for Shallow Circuits," Electronic Proceedings in Theoretical Computer Science," 2020, vol. 318, pp. 214-229, 10.4204/EPTCS.318.13. 

\bibitem{farhi} E. Farhi, J. Goldstone, S. Gutmann, ``A Quantum Approximate Optimization Algorithm'', 2014, \textit{arXiv:1411.4028}.
\bibitem{googleai} M. Harrigan et al, "Quantum Approximate Optimization of Non-Planar Graph Problems on a Planar Superconducting Processor," 2021, Nature Physics 17, pp. 332-336.
\bibitem{wang}Z. Wang, "Quantum approximate optimization algorithm for MaxCut: A fermionic view," 2018, Phys. Rev. A 97, 022304.
\bibitem{wurtz} J. Wurtz and P. Love, "Classically optimal variational quantum algorithms," \textit{	arXiv:2103.17065}.
\bibitem{ruslan}R. Shaydulin and Y. Alexeev, "Evaluating Quantum Approximate Optimization Algorithm: A Case Study," in 2019 Tenth International Green and Sustainable Computing Conference (IGSC), Alexandria, VA, USA, 2019 pp. 1-6, doi: 10.1109/IGSC48788.2019.8957201.

\bibitem{grant} E. Grant, T. Humble, B. Stump, "Benchmarking Quantum Annealing Controls with Portfolio Optimization", PHYSICAL REVIEW APPLIED 15, 014012, 2021.
\bibitem{neukart} F. Neukart et al "Traffic flow optimization using a quantum annealer", Frontiers in ICT, 4:29, 2017.

\bibitem{pesah} A. Pesah et al, "Absence of Barren Plateaus in Quantum Convolutional Neural Networks", Phys. Rev. X 11, 041011, 2021.

\bibitem{seyon} Seyon Sivarajah et al, "t$|$ket$\rangle$: a retargetable compiler for NISQ devices," 2021, Quantum Sci. Technol. 6 014003.

\end{thebibliography}
\end{document}